\PassOptionsToPackage{dvipsnames}{xcolor}
\documentclass[10pt,sigconf,nonacm,screen]{acmart}

\usepackage[all]{nowidow}
\usepackage{xcolor}
\usepackage{subcaption}
\usepackage{hyperref}
\usepackage{acronym}
\usepackage{pdfpages}

\usepackage[capitalise]{cleveref}
\crefformat{section}{\S#2#1#3} 
\crefformat{subsection}{\S#2#1#3}
\crefformat{subsubsection}{\S#2#1#3}

\begin{document}
\includepdf[pages=-, scale=1.0]{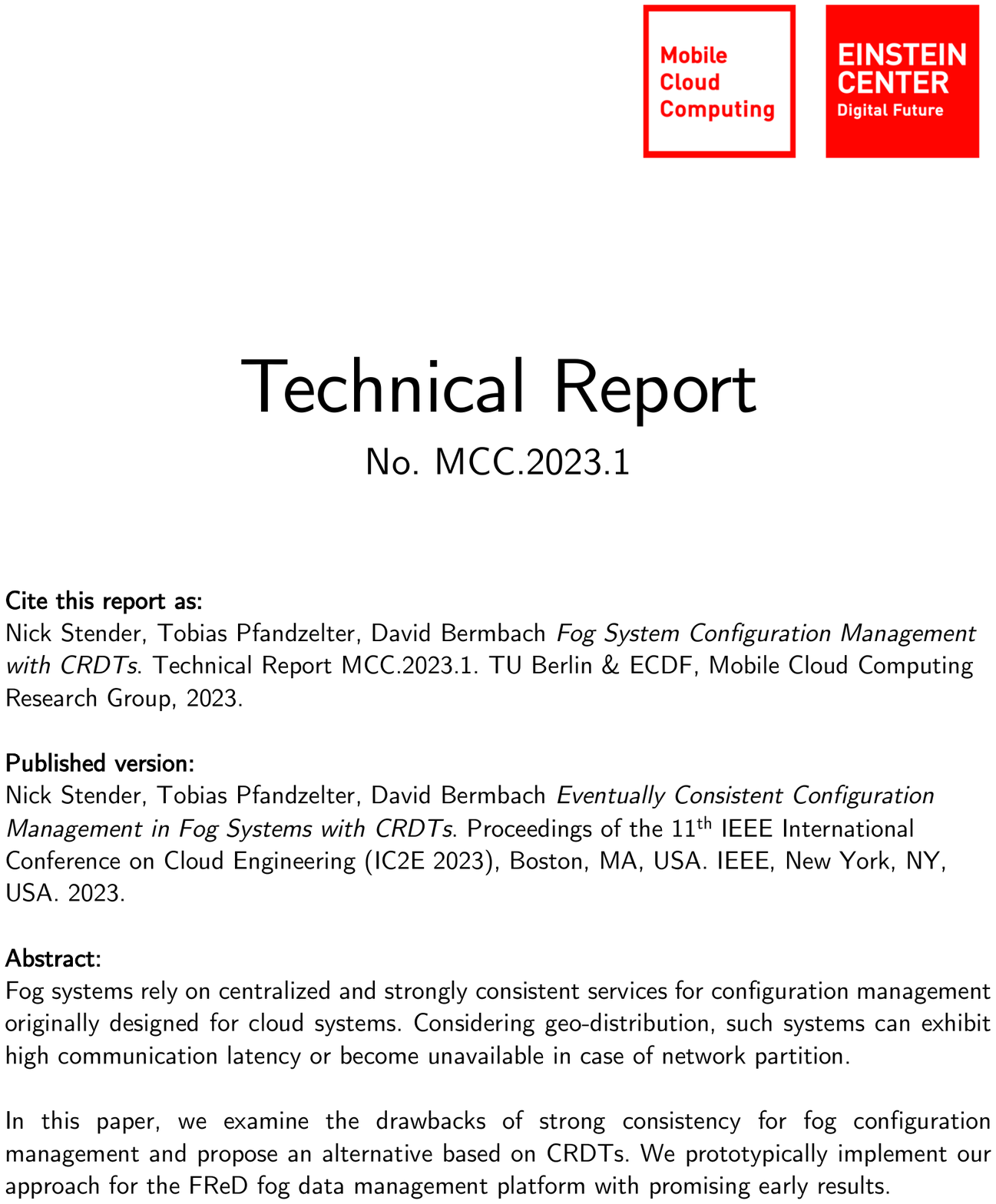}

\author{Nick Stender}
\affiliation{%
    \institution{TU Berlin \& ECDF}
    \city{Berlin}
    \country{Germany}}
\email{nst@mcc.tu-berlin.de}

\author{Tobias Pfandzelter}
\affiliation{%
    \institution{TU Berlin \& ECDF}
    \city{Berlin}
    \country{Germany}}
\email{tp@mcc.tu-berlin.de}

\author{David Bermbach}
\affiliation{%
    \institution{TU Berlin \& ECDF}
    \city{Berlin}
    \country{Germany}}
\email{db@mcc.tu-berlin.de}

\title{Fog System Configuration Management with CRDTs}

\begin{abstract}
    Current fog systems rely on centralized and strongly consistent services for configuration management originally designed for cloud systems.
    In the geo-distributed fog, such systems can exhibit high communication latency or become unavailable in case of network partition.
    In this paper, we examine the drawbacks of strong consistency for fog configuration management and propose an alternative based on CRDTs.
    We prototypically implement our approach for the FReD fog data management platform.
    Early results show reductions of server response times of up to 50\%.
\end{abstract}

\maketitle

\section{Introduction}
\label{sec:introduction}

Fog computing combines geo-distributed servers at the edge, in the cloud, and in the core network to support novel application domains such as the IoT and autonomous driving~\cite{bonomi_fog_nodate,yannuzzi_key_2014,paper_bermbach2017_fog_vision,yi_survey_2015}.
Fog platforms, e.g., \emph{FogStore}~\cite{gupta_fogstore_2018} and \emph{FReD}~\cite{pfandzelter2023fred,poster_hasenburg2020_towards_fbase,techreport_hasenburg2019_fbase}, use centralized configuration management systems with strong consistency.
While desirable for easier configuration of replicas and availability clusters, this comes with an inherent performance penalty~\cite{poster_pfandzelter2022_coordination_middleware,vogels2009eventually} that is exacerbated in fog systems, which are highly geo-distributed with connections over the unreliable Internet~\cite{caiza2020fog}.

Eventual consistency could enable distributed configuration management with low latency and highly available access to global configuration data~\cite{poster_pfandzelter2022_coordination_middleware}.
In this paper, we explore the potential QoS benefits of such an approach and show the drawbacks of eventual consistency in fog configuration management.
Specifically, we develop an alternative distributed configuration management system with eventual consistency for the fog data management platform FReD.
We convert existing methods and data fields in the configuration management service to use conflict-free replicated data types (CRDTs) that allow resolving consistency conflicts after they occur due to network partitions or delay~\cite{shapiro2011comprehensive,jeffery_rearchitecting_2021}.

We make the following contributions:

\begin{itemize}
    \item We design an eventually consistent configuration management architecture for the FReD fog data management platform based on CRDTs (\cref{sec:design}).
    \item We prototypically implement this design and evaluate it experimentally (\cref{sec:evaluation}).
\end{itemize}

\section{Background}
\label{sec:background}

Before we introduce the specific architecture of our system, we will give some background information about the technologies and theoretical concept used in this paper.

\subsubsection*{Fog Computing}

Fog computing extends cloud computing past the confines of a centralized data center by incorporating compute and storage resources in the core network and the edge~\cite{paper_bermbach2017_fog_vision,bonomi_fog_nodate,osanaiye_cloud_2017}.
Fog systems are deployed across geo-distributed heterogeneous nodes close to end users and devices in order to provide application services with low latency, decrease network strain, and increase data protection.

Managing applications within such an environment is more complex than in the cloud given heterogeneity and geo-distribution.
Researchers have proposed compute~\cite{paper_pfandzelter2020_tinyfaas}, messaging~\cite{paper_hasenburg2020_disgb}, and data management~\cite{gupta_fogstore_2018,poster_hasenburg2020_towards_fbase,techreport_hasenburg2019_fbase,pfandzelter2023fred} abstractions to make adopting fog computing easier.

The FReD fog data management platform~\cite{pfandzelter2023fred} provides the abstraction of \emph{keygroups}, logically coherent data tables that can be accessed by application with a key/value interface.
For each keygroup applications can specify geographically diverse replication locations.
As a central source of truth about the available FReD locations, replication instructions, and user authentication, the FReD \emph{naming service} runs a centralized \texttt{etcd}~\cite{etcd} cluster in the cloud.
This is a similar design to \emph{Apache ZooKeeper}~\cite{hunt_zookeeper_nodate}.

\subsubsection*{Consistency in Distributed Systems}

In distributed systems, there are different levels of data consistency.
In strong consistency, two copies of a data item are identical at all times of valid system state~\cite{terry_replicated_2013}.
Eventual consistency describes the promise that data will be identical or consistent at some point in the future.
In strong eventual consistency two data copies that receive the same updates, albeit not necessarily in the same order, will end up in the same state eventually~\cite{shapiro2011comprehensive}.

The PACELC theorem describes that distributed computing systems have to choose between consistency and latency in normal operation.
In case of a network partition, a choice between consistency and availability has to be made~\cite{abadi2012consistency,golab2018proving}.
In a cloud context, most strongly consistent configuration management systems choose consistency in both cases as low communication latency and high network availability can be assumed in a data center.
In fog computing, which applications use to decrease communication latency and to rely less on unstable Internet connections, emphasis should instead be put on latency in normal operation and availability during network partitions~\cite{paper_bermbach2017_fog_vision,madsen_reliability_2013}.

\subsubsection*{CRDTs}

Conflict-free replicated data types come in \emph{state-based} and \emph{operation-based} variants~\cite{shapiro2011comprehensive}.
While operation-based CRDTs are more efficient in communication, they require \emph{exactly-once} message delivery.
We thus focus on state-based CRDTs that are more compatible with gossip dissemination in distributed fog systems~\cite{baquero_making_nodate}.

A state-based CRDT is a tuple of data type and merge function.
This function takes two data items and produces a combined output item, so that the states of two nodes can be combined without communication between them.
A \emph{Last-Write-Wins element set} (LWW) is a state-based CRDT based on an \emph{add} and a \emph{remove} set~\cite{ahuja_edge_2019,shapiro2011comprehensive}.
Inserted elements are added to the \emph{add} set, and removed elements are added to the \emph{remove} set.
Both set additions include timestamps.
An element is considered an element of the LWW if a) it is only present in the \emph{add} set \textbf{OR} b) present in both sets, but the timestamp of the entry in the \emph{add} set is newer.
Unique replica identifiers may be added to per-replica counters to ensure timestamp uniqueness.

\section{CRDT-Based Configuration Management in FReD}
\label{sec:design}

We propose replacing the centralized \texttt{etcd} naming service in FReD with a decentralized CRDT-based approach with the goal of improving client access latency and network partition tolerance.
This is especially relevant as reading configuration data is on the hot path of a client request to FReD:
When a client reads data from a FReD node, the node has to check that the client is allowed to perform this read.
Similarly, when an update request occurs, the FReD node has to read keygroup configuration that specifies to which other nodes in the fog network data should be replicated.

We use LWW element sets to hold configuration data in our eventually consistent configuration service.
Specifically, we use one set each for node information, keygroup configuration, system permission, and FReD node organization.
As there is no longer any central instance, we use a distributed bootstrapping approach where new nodes are informed of one existing node to create a decentralized overlay network.
We use a gossip-style message dissemination where nodes periodically call other nodes to update their view of the network and discover unavailable nodes~\cite{birman_promise_2007}.
We convert the following functionality of the FReD naming service:

\emph{Node Registration:}
Instead of registering a new node with a central orchestrator, node identifier and address are sent to the providing bootstrapping node.
As node creation happens infrequently and identifiers can easily be made unique, this is unlikely to lead to incorrect behavior.
In case of a restart after failure LWW ensures that outdated information about a node is overwritten.

\emph{User Permission Changes:}
When an administrator makes permission changes for a user at the user's node, this node will immediately apply those changes.
If message dissemination is slower than user movement, data staleness could lead to user permissions being outdated when switching nodes.
The correlation between physical locations of users and nodes, and the data dissemination latency, however, makes this unlikely.
Partitioned nodes are a challenge, as updated permission information cannot reach them.
The only alternative to stale information is unavailability of the node, e.g., by disabling access for users when the partition is detected.

\emph{Keygroup Modification:}
When creating keygroups, identifier uniqueness is paramount.
Concurrent creation of two keygroups with identical names at two different keygroups will lead to conflicts in LWW.
However, the large identifier space makes this unlikely.

\emph{Keygroup Membership:}
Administrators and application can join and remove nodes from keygroups to specify data replication.
Conflicts in an eventually consistent configuration management could occur only for changes made to the same keygroup, as memberships to keygroups is independent.
If keygroup membership for a single node is modified concurrently, one of these changes is overwritten by LWW.
Such a situation is unlikely, however, as, logically, each keygroup is managed by a single application.

\section{Evaluation}
\label{sec:evaluation}

\begin{figure}
    \centering
    \includegraphics[width=1.0\linewidth]{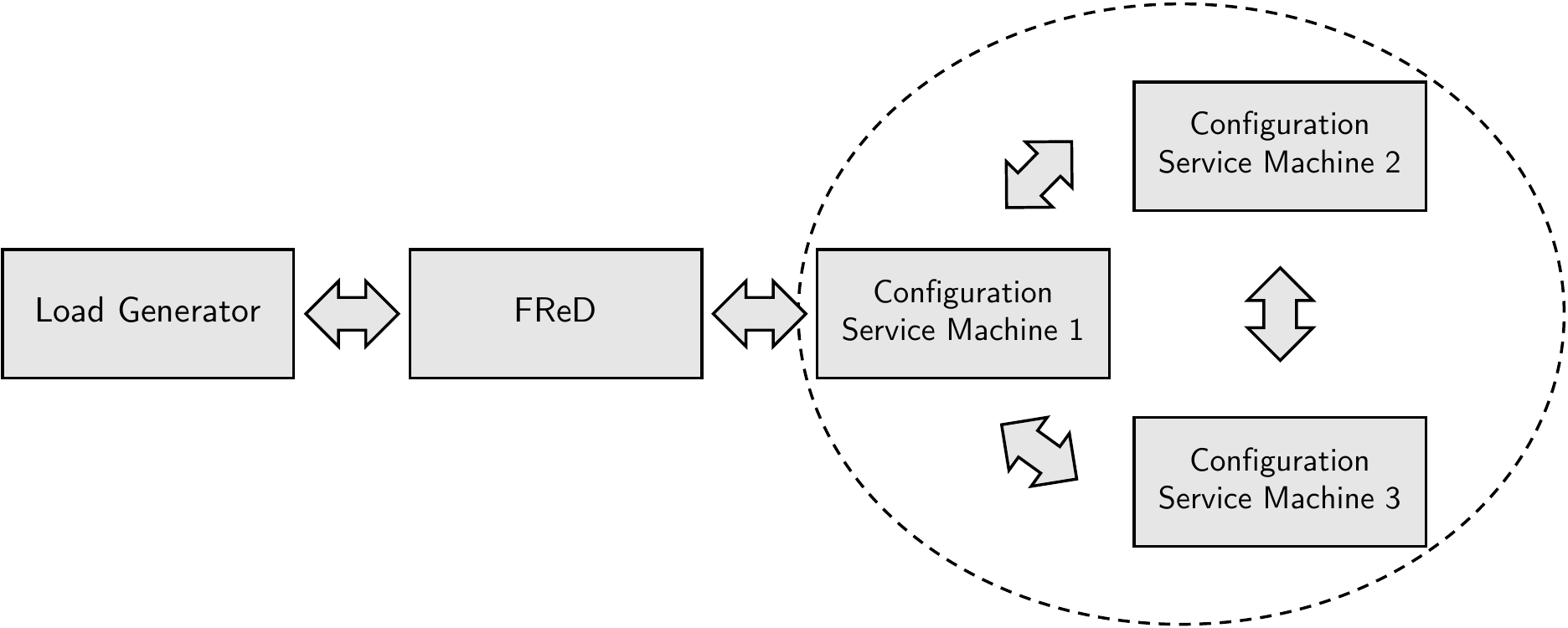}
    \caption{Overview of the different components.}
    \label{fig:experimentSetup}
\end{figure}

We implement the alternative naming service using \emph{Go}~\cite{go} and \emph{gRPC}~\cite{grpc}, making it compatible with the open-source FReD implementation.
In our experiments, we start FReD nodes as Docker containers and connect them to a FReD naming service, either the original \texttt{etcd} implementation or our new CRDT-based system.
Each naming service is distributed over at least three machines.
We inject an artificial network delay between containers using \texttt{tc-netem}~\cite{brown2006traffic}.
We connect a load generator to a FReD node that measures completion times of requests.
Our experiment topology is shown in \cref{fig:experimentSetup}.

\subsubsection*{Baseline}
\label{sec:eval:raw}

\begin{figure}
    \centering
    \begin{subfigure}{0.49\columnwidth}
        \centering
        \includegraphics[width=1.0\linewidth]{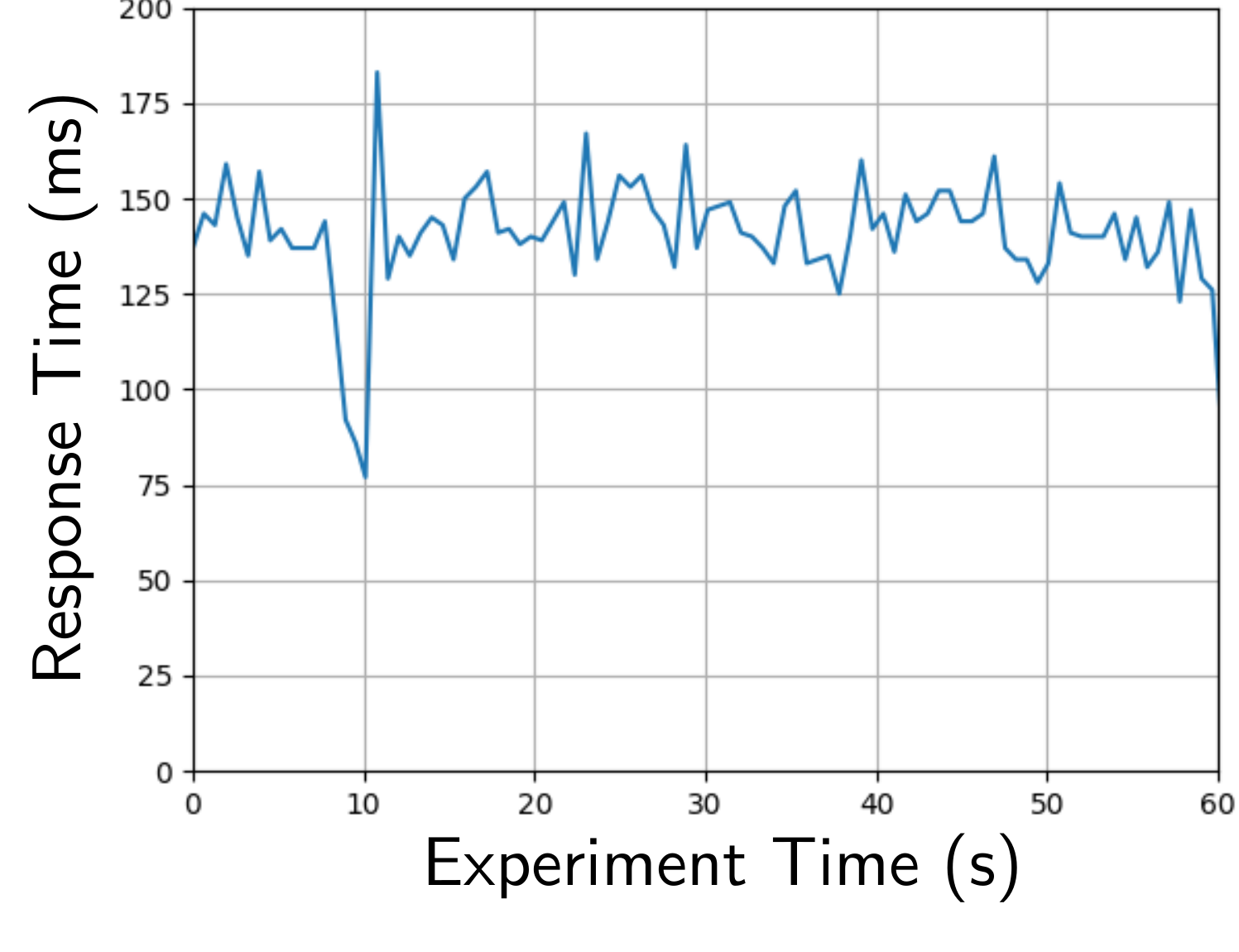}
        \caption{\texttt{etcd}}
        \label{fig:baseline:etcd}
    \end{subfigure}
    \hfill
    \begin{subfigure}{0.49\columnwidth}
        \centering
        \includegraphics[width=1.0\linewidth]{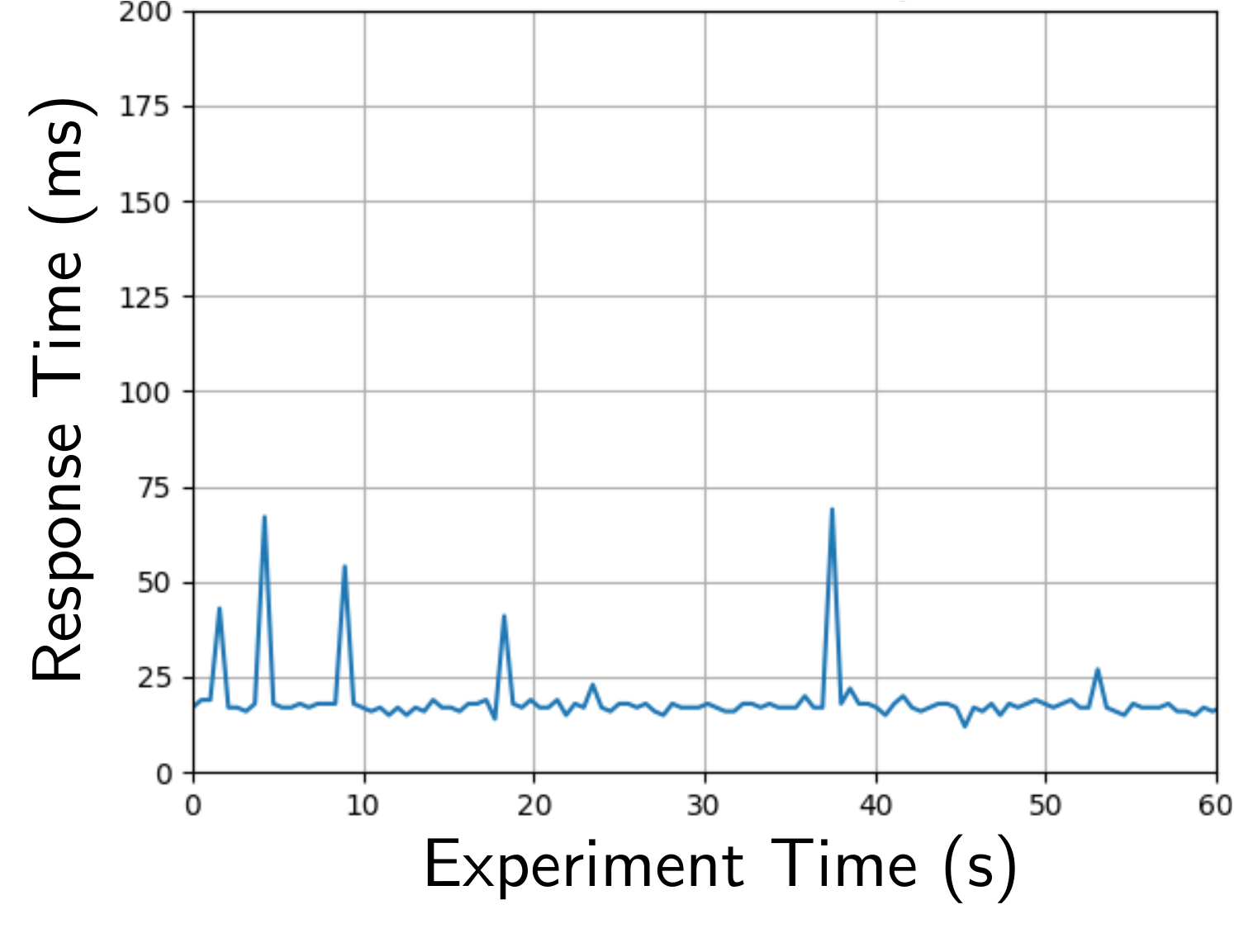}
        \caption{CRDT}
        \label{fig:baseline:crdt}
    \end{subfigure}
    \caption{FReD response times without network delays between configuration service machines using \texttt{etcd} (\cref{fig:baseline:etcd}) and CRDTs (\cref{fig:baseline:crdt}).}
    \label{fig:baseline}
\end{figure}

As a baseline, we compare configuration management approaches without network delay.
To invoke write access to the naming service call the \texttt{createKeygroup} API of FReD to create keygroups from our load generator.
The results in \cref{fig:baseline} show a higher delay for the \texttt{etcd} naming service.
Although we expect this improvement to be caused mainly by the switch to a CRDT-based approach, we cannot rule out that our prototypical implementation is otherwise more efficient than production-ready \texttt{etcd}.

\begin{figure}
    \centering
    \includegraphics[width=0.6\linewidth]{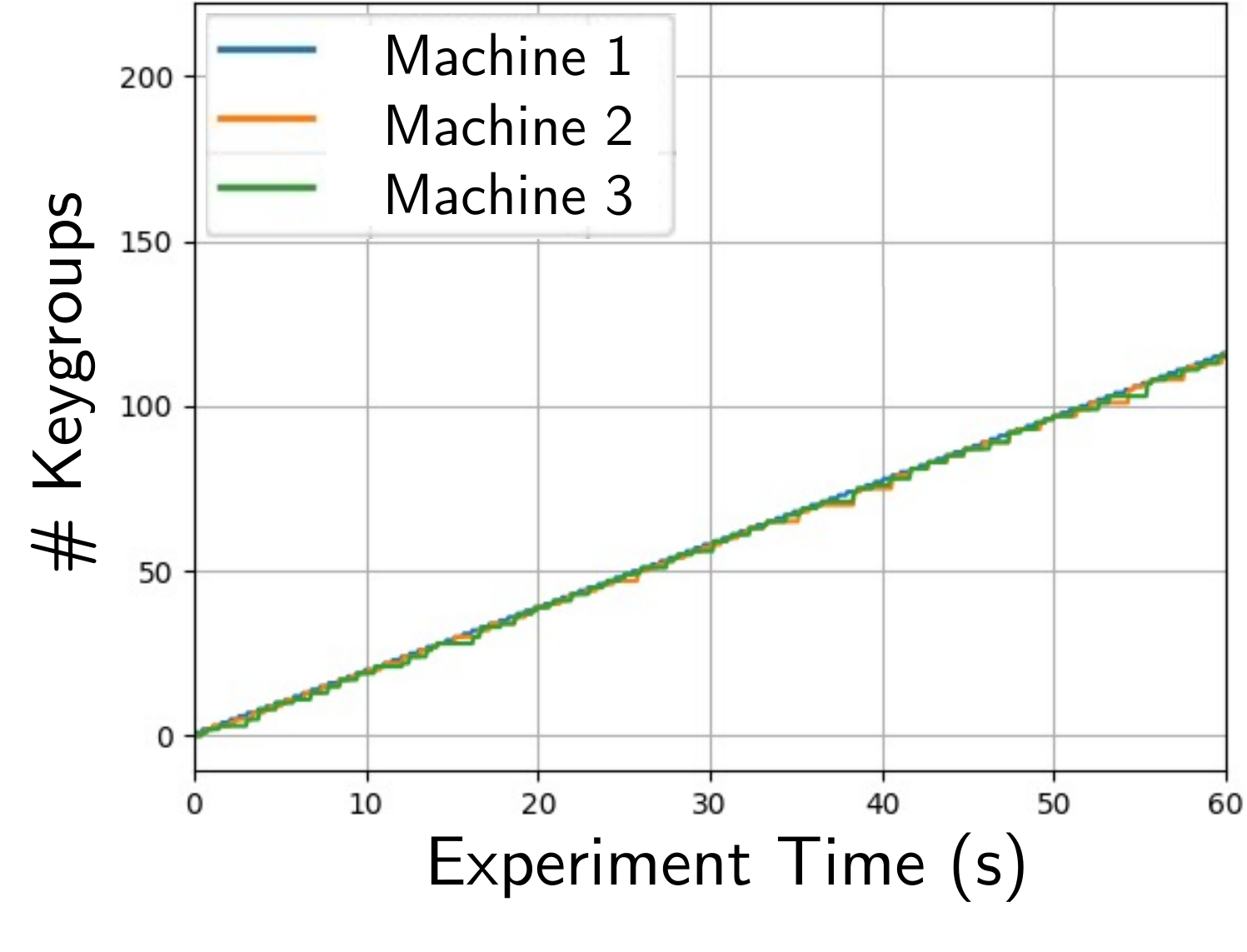}
    \caption{Number of keygroups present in each CRDT-based configuration service instance without added network delays.}
    \label{fig:baseline-number}
\end{figure}

We measure the message dissemination delay in the CRDT-based system by logging the number of keygroups each configuration machine knows about.
As shown in \cref{fig:baseline-number}, the distributed CRDT-based systems converges quickly.

\subsubsection*{With Network Delays}

\begin{figure}
    \centering
    \begin{subfigure}{0.49\columnwidth}
        \centering
        \includegraphics[width=1.0\linewidth]{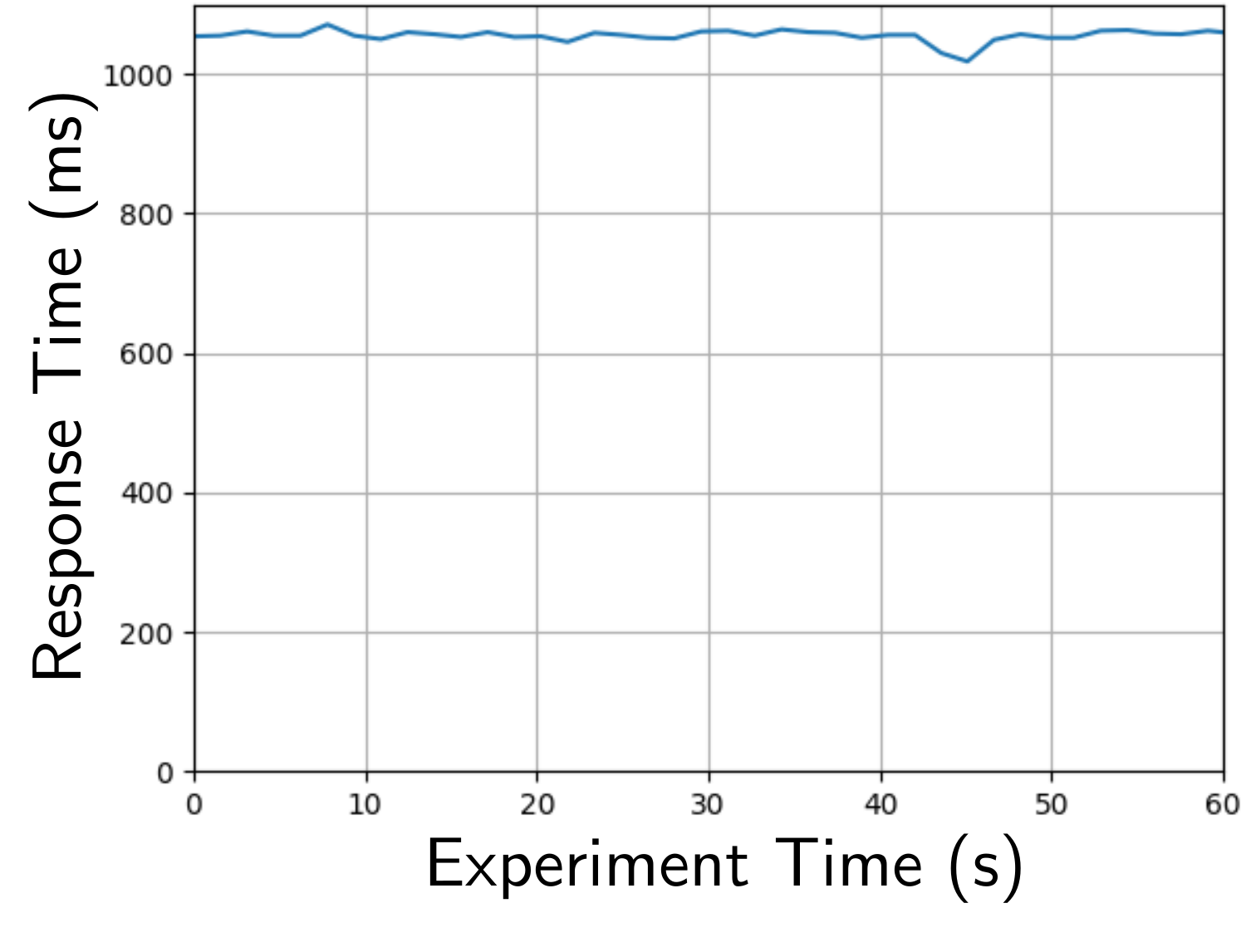}
        \caption{\texttt{etcd}}
        \label{fig:delay:etcd}
    \end{subfigure}
    \hfill
    \begin{subfigure}{0.49\columnwidth}
        \centering
        \includegraphics[width=1.0\linewidth]{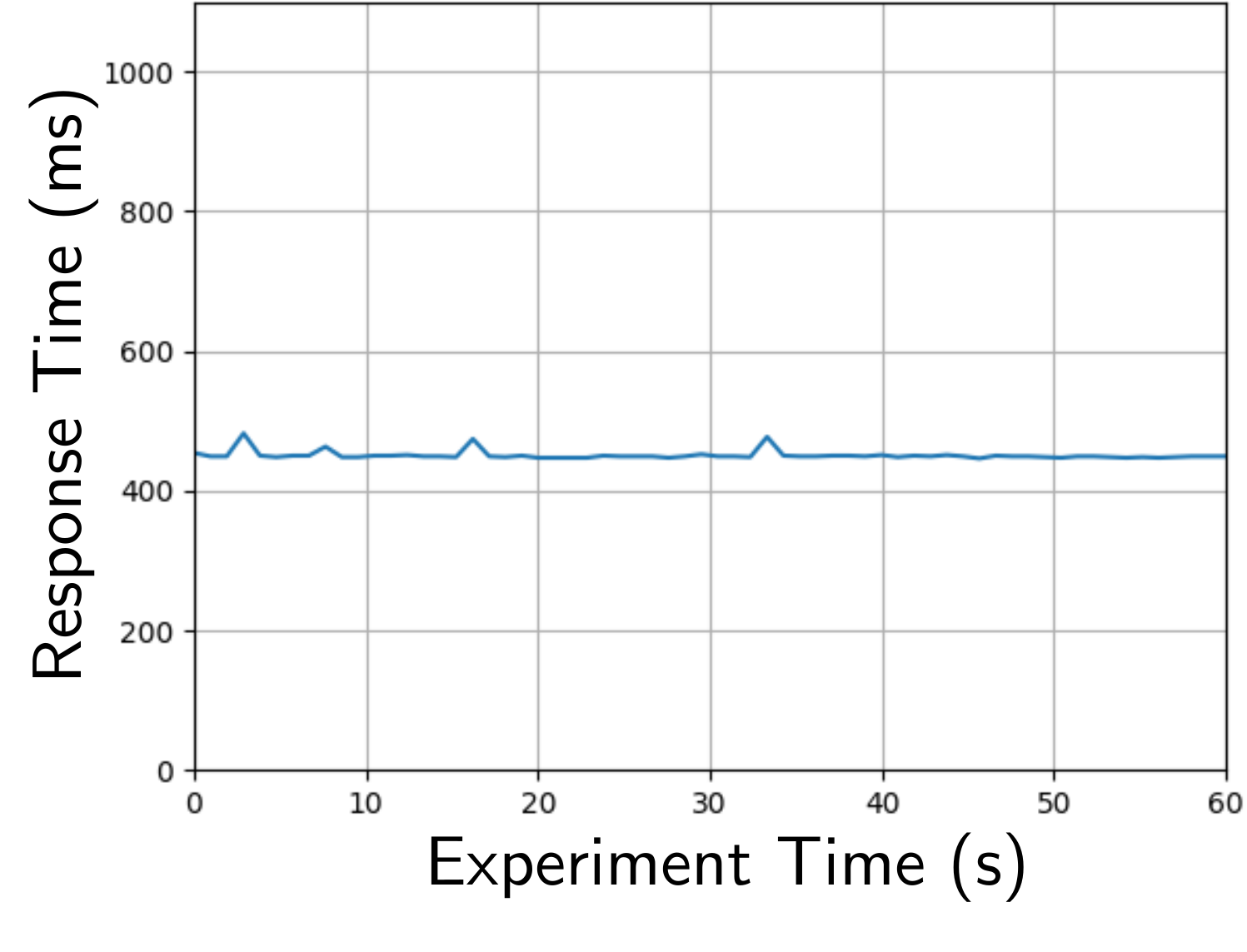}
        \caption{CRDT}
        \label{fig:delay:crdt}
    \end{subfigure}
    \caption{FReD response times with 10ms network delays between configuration service machines using \texttt{etcd} (\cref{fig:delay:etcd}) and CRDTs (\cref{fig:delay:crdt}).}
    \label{fig:delay}
\end{figure}

Using an artificial network delay of 10ms, we evaluate the impact of communication delay between naming service machines.
As shown in \cref{fig:delay}, this small communication delay increases FReD response times for both implementations.
However, the total impact is more noticeable for the \texttt{etcd} naming service.

\begin{figure}
    \centering
    \includegraphics[width=0.6\linewidth]{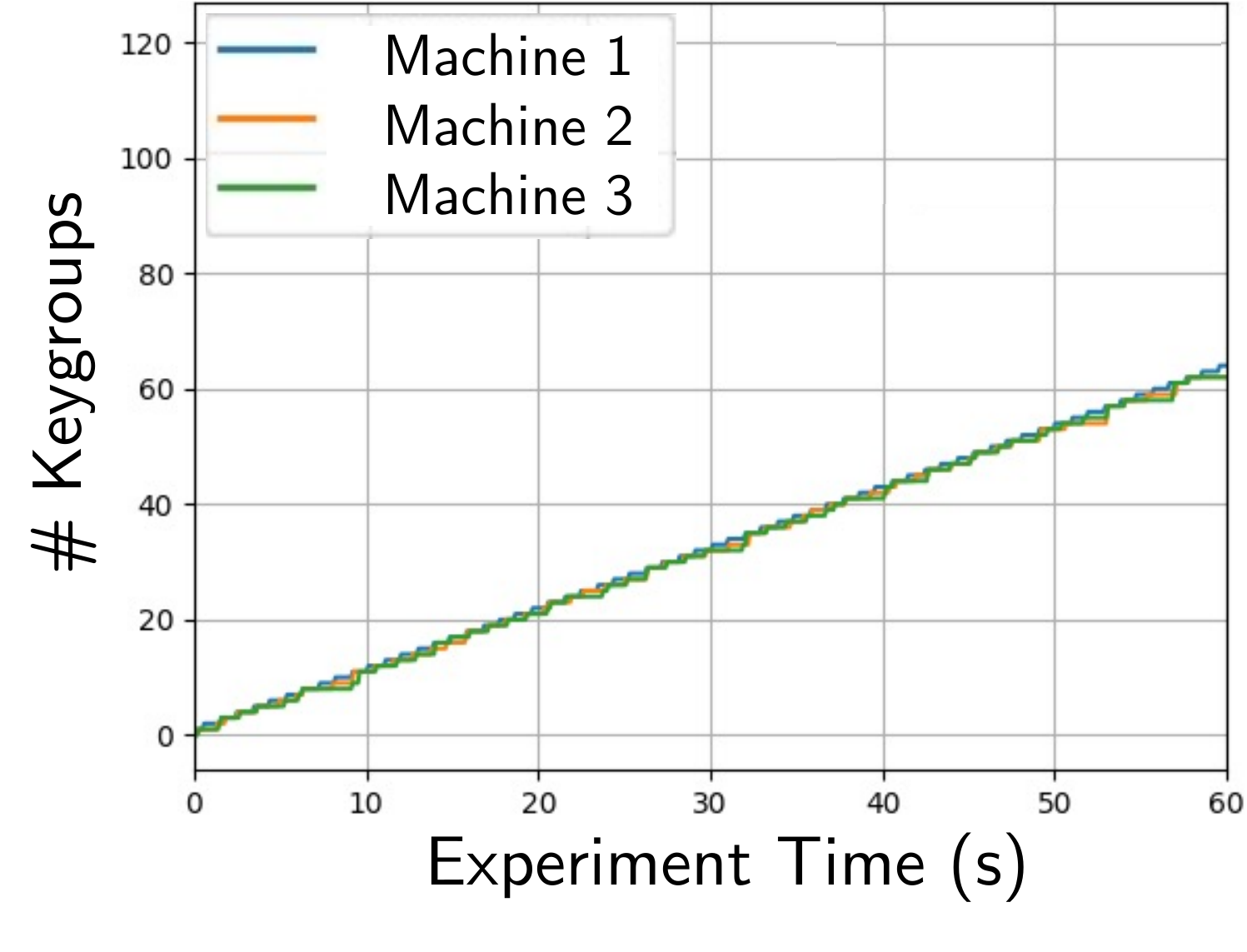}
    \caption{Number of keygroups present in each CRDT-based configuration service instance with 10ms added network delays.}
    \label{fig:delay-number}
\end{figure}

As shown in \cref{fig:delay-number}, there is a slight impact to data dissemination in our CRDT-based configuration management service.

\subsubsection*{Network Partitions}

\begin{figure}
    \centering
    \begin{subfigure}{0.49\columnwidth}
        \centering
        \includegraphics[width=1.0\linewidth]{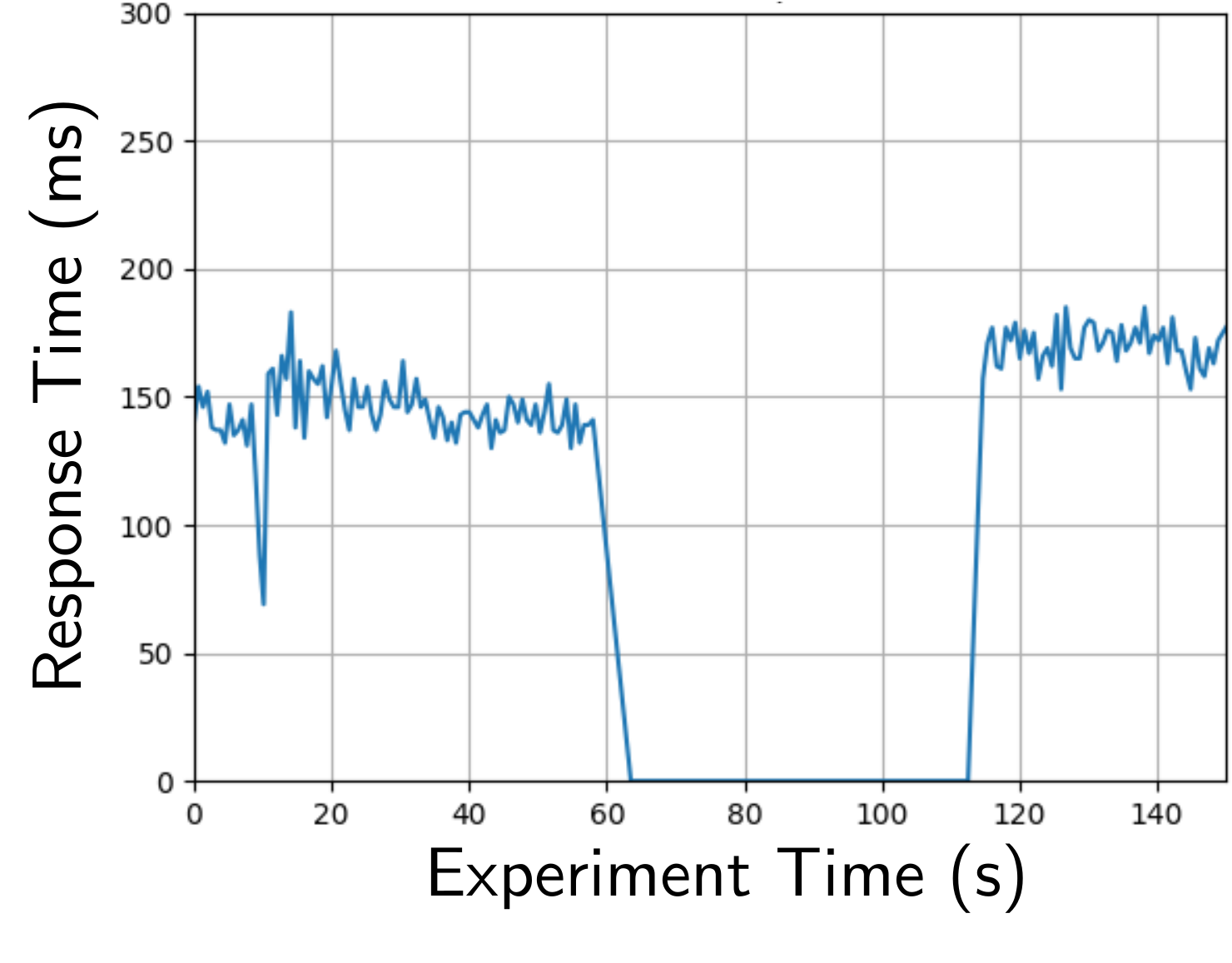}
        \caption{\texttt{etcd}}
        \label{fig:partition:etcd}
    \end{subfigure}
    \hfill
    \begin{subfigure}{0.49\columnwidth}
        \centering
        \includegraphics[width=1.0\linewidth]{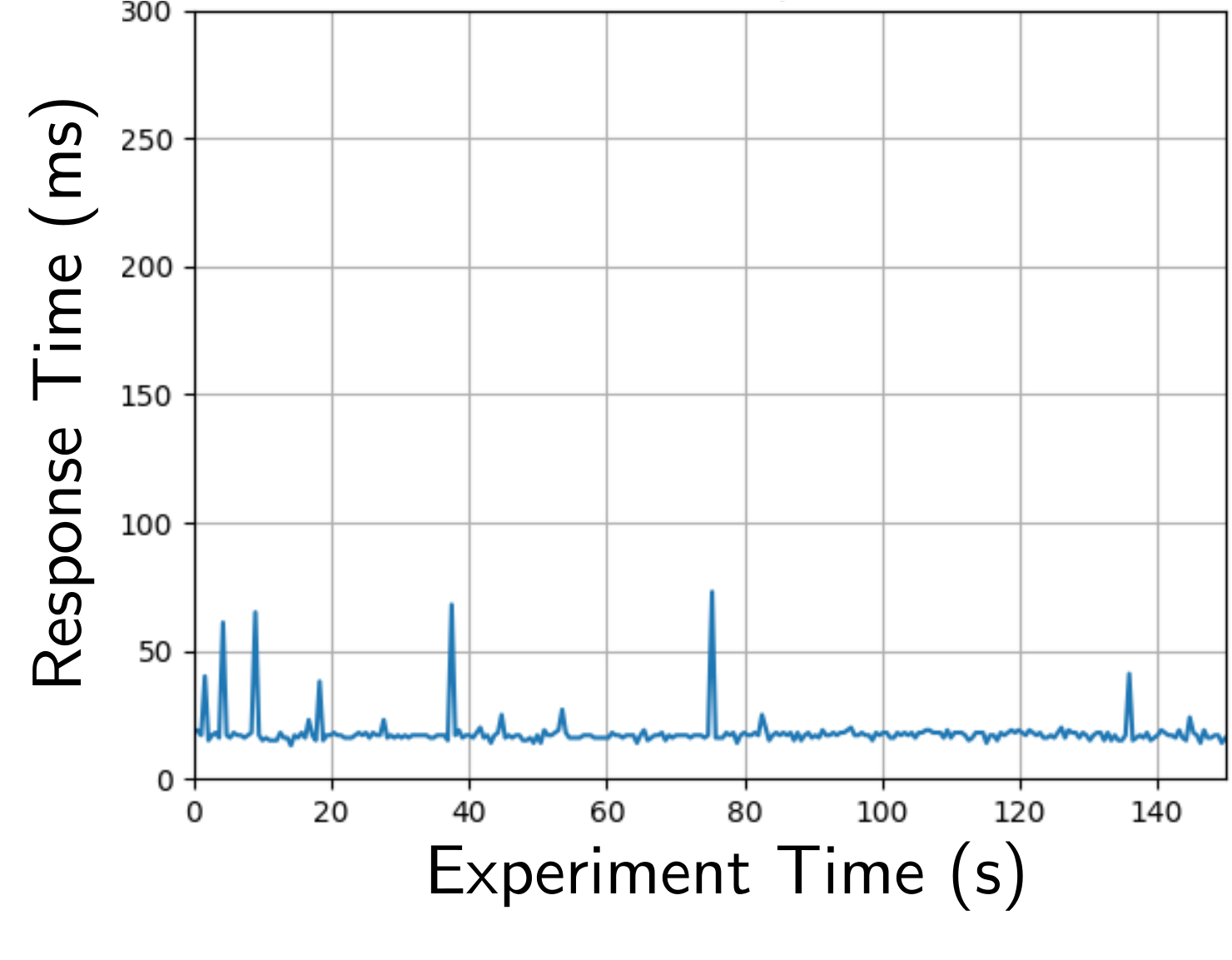}
        \caption{CRDT}
        \label{fig:partition:crdt}
    \end{subfigure}
    \caption{FReD response times with configuration service machines using \texttt{etcd} (\cref{fig:partition:etcd}) and CRDTs (\cref{fig:partition:crdt}). After 45s experiment time, we partition the network between configuration service machines.}
    \label{fig:partition}
\end{figure}

Finally, we introduce a network partition between the naming service machine used by our FReD node and the two others.
This partition is introduced after running the experiment for 45 seconds.
We re-enable the network link after a further 35 seconds.
As shown in \cref{fig:partition}, the partition impacts only the strongly consistent \texttt{etcd} implementation, where all requests fail during the partition (shown as a 0ms response time).
Note also that it takes an additional 20 seconds after the network links are re-enabled for the system to recover.
The CRDT-based implementation remains unaffected by this partition.

\begin{figure}
    \centering
    \includegraphics[width=0.6\linewidth]{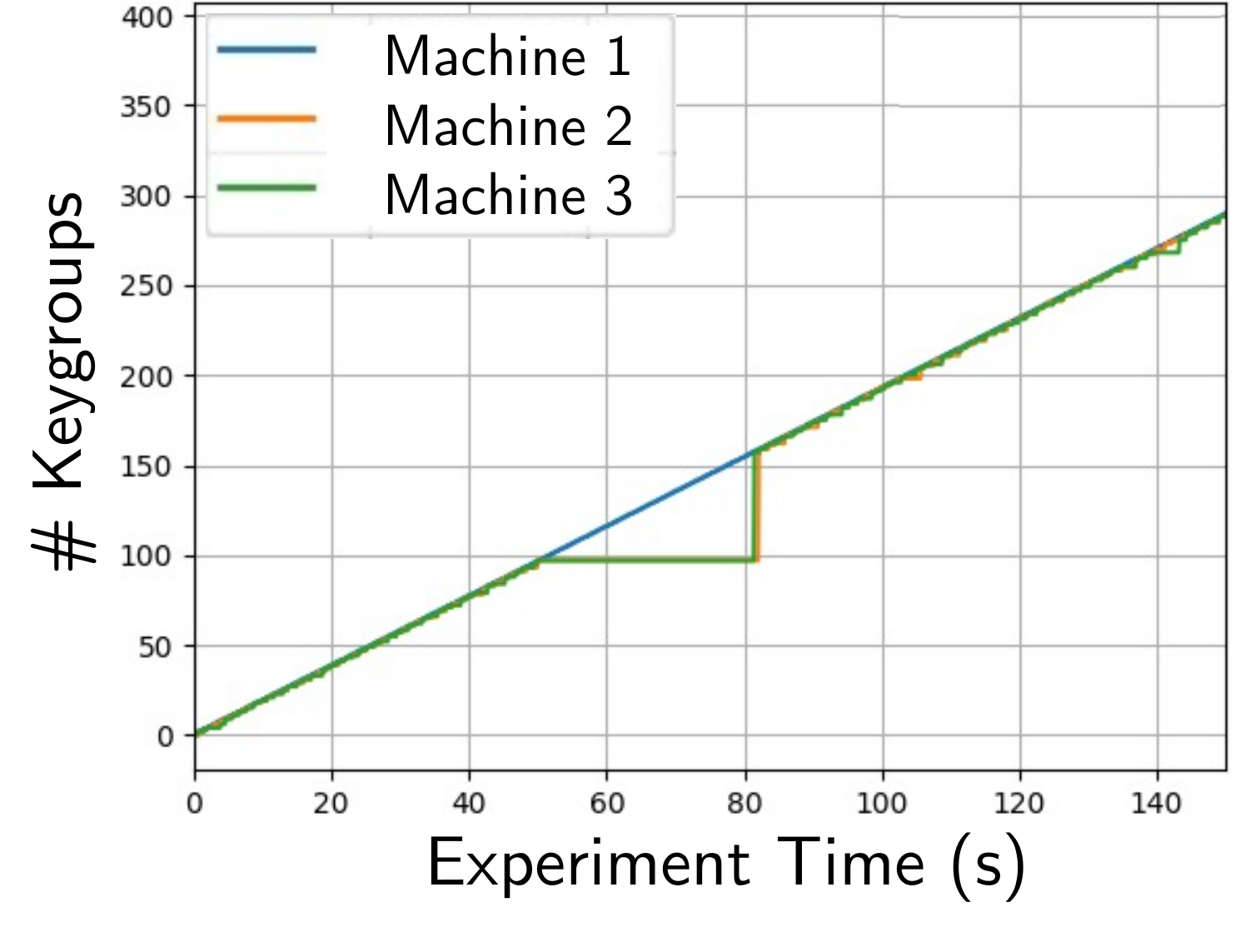}
    \caption{Number of keygroups present in each CRDT-based configuration service instance with network partition after 45s experiment time.}
    \label{fig:partition-number}
\end{figure}

The partition still impacts data dissemination as \cref{fig:partition-number} shows:
During the partition, the other two machines of the naming service do not receive updated keygroup information.
As soon as the network connection is reinstated data is updated again.

\section{Related Work}
\label{sec:related}

To the best of our knowledge, we are the first to implement a CRDT-based configuration management for fog systems.
We first suggested such an approach in prior work~\cite{poster_pfandzelter2022_coordination_middleware}.
In related domains, F{\"o}rd{\H{o}}s and Cesarini~\cite{fordos_crdts_2016} propose CRDT-based configuration management for distributed Erlang systems.
They find their approach to improve response times and system reliability.
Jeffery et al.~\cite{jeffery_rearchitecting_2021} outline a CRDT-based replacement for \texttt{etcd} in distributed Kubernetes.
Although they do not provide experimental evaluation of this approach, this proves the general idea we follow in this paper.
\emph{Serf}~\cite{serf} is a distributed cluster management tool based on the \emph{SWIM} gossip protocol~\cite{das_swim_2002}.
\emph{Serf} is decentralized, available during network partitions, and provides weak consistency guarantees.
It primarily targeting cloud and cluster deployments and its applicability to geo-distributed fog computing system is unclear.

\section{Conclusion \& Future Work}
\label{sec:conclusion}

In this paper, we have shown that eventually consistent configuration management systems based on CRDTs are a promising alternative to strictly consistent centralized solutions for fog systems.
Our evaluation of a CRDT-based distributed naming service for the FReD fog data management platform has shown reduced response times for clients, especially with network delay between machines.
Future work will include a more comprehensive evaluation of the drawbacks of using eventually consistent configuration management in the fog.
We also plan to explore the combination of strong consistency for some configuration data and eventual consistency for others.
While complex, such a hybrid approach would allow for more efficient data dissemination without impacting application logic.

\begin{acks}
    Supported by the \grantsponsor{DFG}{Deutsche Forschungsgemeinschaft (DFG, German Research Foundation)}{https://www.dfg.de/en/} -- \grantnum{DFG}{415899119}.
\end{acks}

\balance
\bibliographystyle{ACM-Reference-Format}
\bibliography{bibliography.bib}

\end{document}